\documentclass[onefignum,onetabnum]{siamonline220329}
\usepackage{amsmath,amsfonts}
\usepackage{array}
\usepackage{algorithmic}
\usepackage{algorithm}
\usepackage[caption=false,font=normalsize,labelfont=sf,textfont=sf]{subfig}
\usepackage{textcomp}
\usepackage{stfloats}
\usepackage{url}
\usepackage{verbatim}
\usepackage{graphicx}
\usepackage{cite}
\usepackage{adjustbox}
\usepackage{multirow}
\usepackage{amssymb}
\usepackage{pifont}
\newcommand{\cmark}{\ding{51}}%

\DeclareMathOperator*{\argmin}{arg\,min}

\begin{document}

\title{Adversarial Generative NMF for Single Channel Source Separation}

\author{{Martin Ludvigsen and Markus Grasmair}\\
  \medskip
  Department of Mathematical Sciences,\\ NTNU -- Norwegian University of Science and Technology,\\ 7491 Trondheim, Norway}

\maketitle

\begin{abstract}
  The idea of adversarial learning of regularization functionals
  has recently been introduced in the wider context of inverse problems.
  The intuition behind this method is the realization that
  it is not only necessary to learn the basic features that make up a class of signals
  one wants to represent, but also, or even more so, which features to avoid in the representation.
  In this paper, we will apply this approach to the
  problem of source separation by means of non-negative matrix factorization (NMF)
  and present a new method for the adversarial training of NMF bases.
  We show in numerical experiments, both for image and audio separation,
  that this leads to a clear improvement
  of the reconstructed signals, in particular in the case where little or
  no strong supervision data is available.
\end{abstract}


\section{Introduction}
Single Channel Source Separation (SCSS) is a type of inverse problem that is concerned with the recovery of individual source signals from a measured mixed signal.
This problem arises in various real-world applications, such as speech and music processing, biomedical signal analysis, and image processing. 
In such applications, the observed signal can be modelled as a linear combination of multiple sources, and the objective is to estimate the underlying sources from this mixture.
For this purpose, generative data-driven approaches have proven to be widely applicable, in particular methods based on Non-Negative Matrix Factorization (NMF).

Two aspects of utilizing NMF for source separation problems have not been discussed much in literature, though, namely how to utilize all available
data for training NMF bases, as well as how to tune parameters for specific source separation problems.
Generative methods like NMF can be trained independently of the specific problem they are applied to,
and thus they have a wide range of applicability.
This makes them more generalizable compared to discriminative methods like end-to-end neural networks
that must be trained for specific problems. 
However, this flexibility comes at the cost of performance and efficiency for specific problems.

In this paper, we will propose a new NMF method for SCSS
that is based on the idea of adversarial regularization, previously
introduced for more general inverse problems.
Our contribution mainly consists of:
\begin{itemize}
\item An extension of existing work on adversarial regularization functions
  to generative models and specifically models for source separation problems. 
\item A particular parameterization of adversarial regularization functions
  that leads to what we call \emph{Adversarial Generative Non-Negative Matrix Factorization} (ANMF).
\item A numerical algorithm for training the proposed methods.
\end{itemize}

Our method leverages mixed data and data from other sources during training
and can be utilized in weak supervision settings.
We will also demonstrate how our method can be integrated into a flexible
framework for fitting NMF bases that utilize both weak and strong
supervision data.
In Section~\ref{sec:background}, we will provide an overview over the problem
of single channel source separation and NMF.
Then we will introduce our main ideas and methods for the adversarial
generation of NMF bases in~\ref{sec:adversarial}.
In Section~\ref{sec:numerics}, we will discuss the numerical implementation
using a multiplicative algorithm.
Moreover, we will construct stochastic variants of that algorithm.
Finally,
we will show in Section~\ref{sec:numerical} by means of numerical experiments,
both for image and audio data,
that the proposed method yields better quantitative and qualitative results
than existing NMF-based methods for source separation problems.

\section{Background}
\label{sec:background}

\subsection{Single Channel Source Separation}
The problem of single channel source separation in finite dimensions consists of recovering the
$S$ individual sources $u_i \in \mathbb{R}^m$ and potentially also the weights $0 \le a_i\le 1$,
$\sum_{i = 1}^S a_i = 1$, from a measured mix $v \in \mathbb{R}^m$, modelled as
\begin{equation}
  \sum_{i = 1}^S a_i u_i = A\mathbf{u} = v.
  \label{eq:SCSS}
\end{equation}
The term ``single channel'' comes from the fact that we assume to only have
one measurement of $v$, as opposed to the multichannel case where
we have several measurements of the same mix, but with different weights.
The latter is for example the case in audio recordings with more than one microphone.
Depending on the problem setting, we may be only interested in the recovered signals
$\tilde{u}_i = a_i u_i$, but not in the actual value of the weights.
Also, in some applications only some sources need to be recovered,
whereas others represent noise.

The problem of solving~\eqref{eq:SCSS} is an underdetermined inverse problem.
If the weights $a_i$ are known, we have a linear inverse problem;
else~\eqref{eq:SCSS} can be either seen as a non-linear inverse problem,
or as a linear inverse problem with uncertain operator $A$.

\subsection{Notation and data settings}
Because the problem of solving~\eqref{eq:SCSS} is underdetermined,
we require some prior knowledge about the structure of the different sources $u_i$
and also the weights $a_i$ and the mixed data $v$.
For this, we use a probabilistic framework, where we assume that we have some
information about the probability distribution of the different possible source signals.
This typically comes in the form of samples of the source signals,

For single channel source separation, there are
several data settings of interest
that differ in the type of available data:
\subsubsection*{Strong supervision}
  Here we have access to samples $v^{(k)}$ of the mixed data
  together with the correct separated sources $u_i^{(k)}$
  and the corresponding weights $a_i^{(k)}$
  such that $v^{(k)} = \sum_i a_i^{(k)} u_i^{(k)}$.
  In our probabilistic setting, this means that we know the
  \emph{joint} probability distribution
  $\mathbb{P}_{A \times U \times V}$ of weights, sources, and data.
  This allows for the training of dedicated discriminative models for specific problems,
  which usually yield superior results compared to models trained in weaker supervision settings.
  However, this setting is often unrealistic, as obtaining large amounts of
  strong supervised data can be challenging in practice.

  A variant of this strong supervision setting
  is one where we only have samples from the weighted sources
  $\tilde{u}_i^{(k)} := a_i^{(k)} u_i^{(k)}$, but we do not know
  the correct splitting of $\tilde{u}_i^{(k)}$ into weight and unweighted sample.
  This can be modelled by having access to the joint probability
  distribution $\mathbb{P}_{\tilde{U}\times V}$ of \emph{weighted sources} and mixed data.
\subsubsection*{Weak supervision} This is also called \emph{semi-supervision:}
  There exist different interpretations of the notion of weak supervision, but
  in this paper we will mostly focus on the setting where we
  have access to samples of the different sources $u_i^{(\ell)}$
  and also samples $v^{(k)}$ of mixed data,
  but no information about the correct separation
  $v^{(k)} = \sum_i a_i^{(k)} u_i^{(k)}$.
  In addition, we may have some statistical model for the
  distribution of the weights $a_i$.

  This is a more realistic data setting in many cases, as we can more easily
  obtain additional clean signals of the different sources and mixed signals.
  However, it is much more difficult to obtain mixed signals
  together with the corresponding unmixed signals

  Translated to the probabilistic setting, 
  we have information about the \emph{marginal} distributions
  $\mathbb{P}_{U_i}$ and $\mathbb{P}_V$, and in addition
  some model for $\mathbb{P}_{A_i}$.
  However, we generally cannot assume that the distributions
  are independent, as this would essentially mean that
  each sample from any source can be arbitrarily mixed
  with any sample from a different source.
  Thus, we assume that we have no, or only little, information about
  the joint distribution $\mathbb{P}_{A \times U \times V}$.
  
  A related interesting case is the weaker supervision case
  where we only have access to samples from some sources, but not all.
  This problem is particularly important in denoising applications,
  where we may have access to noisy and clean signals, but no access to samples of clean noise.
\subsubsection*{Synthetic strong supervision}
  This case lies somewhere between strong supervision and weak supervision
  and occurs if we have access to samples of the weighted sources
  $\tilde{u}_i^{(k)} = a_i^{(k)}u_i^{(k)}$ and in addition can
  reasonably assume that their distributions are close to being independent.
  In this case, we can generate synthetic samples of supervised
  data by using the forward model~\eqref{eq:SCSS}
  and thus simulate strong supervision.
\subsubsection*{Blind source separation}
  In this case we only have access to samples of mixed data,
  but not to samples from the unmixed sources.
  This is a hard problem, which usually require
  knowledge-based assumptions on the data, as opposed to using data-driven methods.
  Examples are the famous ROF total variation model for image denoising \cite{rudin1992nonlinear},
  the RPCA model for speech denoising and separating vocals
  from mixed music tracks \cite{huang2012singing},
  and Independent Component Analysis (ICA) \cite{comon1994independent}.
  This problem is more common in multichannel source separation,
  where it is possible to leverage information from multiple recordings of the same signals.
\smallskip

In this paper, we are mainly concerned with weak supervision where creating synthetic supervised data is either infeasible or not of interest.
We propose methods where no joint information is available, but they can just as easily be used in the strong supervision case.
We do this for both the case where data from all sources are available, and later when data from one source is missing.

\subsection{Previous work on source separation}

There are two main branches of data-driven methods for source separation: generative models and discriminative models.

In recent years, purely discriminative models that assume strong supervision have gained a lot of attention, as they often lead to both quantitatively and qualitatively better results
for most problems \cite{luo2022music,hershey2016deep,rouard2022hybrid,8462116}. 
They generally consist of neural networks that either directly learn mappings between mixed and unmixed data.
Some recent works also propose further tuning with semi-supervised data \cite{luo2022music}.
Such models can usually be fine-tuned for specific problems, potentially leading to efficient models. 
This is however also a problem, as it reduces their generalizability.
In these cases, entire models might need to be retrained or tuned for slightly different problem settings, which
is an issue considering the cost of training and storing different models.

Generative approaches, on the other hand, are easier to apply in settings with weaker supervision, as they usually
do not require information about the joint distribution $\mathbb{P}_{\tilde{U} \times V}$,
or even knowledge of the distribution of mixed data $\mathbb{P}_V$. 
Thus, they have a larger potential to utilize information from all available data.
Since they can be used more or less as plug-and-play models,
they tend to generalize better to different problem settings and different data.
The same trained generative model can
be used for different, but related, applications.
For instance, the same generative model might be simultaneously used for 
speech synthesis, speech classification, speech denoising and speech separation.
This can be more efficient than training a separate discriminative
model for each of these tasks.

One traditional approach to generative source separation is Non-Negative Matrix factorization (NMF) \cite{fevotte2018single}.
Recently, it has become more common, though,
to investigate advanced generative models like GANs and autoencoders \cite{subakan2018generative}, deep generative priors \cite{jayaram2020source} and generative latent optimization \cite{halperin2019neural}.
While these more advanced methods can lead to better results, particularly for non-stationary data, NMF is still a method that is easy to understand and implement.
The relatively low complexity of NMF can also make it less vulnerable to adversarial attacks and therefore more robust. 
Another benefit of NMF is its relative stability with respect to different weights, as the NMF generator function
is naturally able to produce signals with different scalings. 
NMF can also be a valuable tool for initializing other more initialization sensitive methods \cite{grais2014deep}.
Therefore we focus on NMF-based methods for most of this article, though many
of our ideas are also applicable to more advanced generative methods.

\subsection{NMF for source separation}
\label{ss:NMFSS}

We will now explain how NMF is usually used for single channel source separation
in the weakly supervised setting:

We assume that we are given training data in the form of samples
$u_i^{(k)} \in \mathbb{R}^m$ for each source $i$.
Our underlying assumption is that these samples can be approximately
written as non-negative linear combinations
$u_i^{(k)} \approx W_i h_i =: g_i(h_i)$ for some yet to be determined basis
$W_i \in \mathbb{R}_+^{m\times d}$ and latent variables $h_i \in \mathbb{R}_+^d$.
Here the dimension $d \ll m$ has to be chosen a-priori and will have
a large effect on the final result, as well as the computational complexity.
It is also possible to choose a different dimension $d_i$ for each of the sources,
which can be important if, for instance, one of the source signals is vastly more
complex than the others.

We note here that, on a more abstract level, we can
interpret the NMF model as the assumption that the
samples lie in the positive cone spanned by
the columns of $W_i$. We will come back to this
interpretation later when we introduce the main novel
ideas of this paper.

\subsubsection*{Training}
  To start with, we collect the samples $u_i^{(k)} \in \mathbb{R}^m$
  for each source $i$ column-wise in a matrix $U_i$.
  For each $i$, we then solve the bi-level problem
  \begin{equation}
    \min_{W_i \ge 0} \|U_i - W_iH(U_i,W_i)\|_F^2 + \mu_W |W_i|_1,\label{eq:NMFtrainingW}
  \end{equation}
  where
  \begin{equation}
    H(U,W) = \argmin_{H \ge 0} \|U_i - W_iH\|_F^2 + \mu_H |H|_1.\label{eq:NMFtrainingH}
  \end{equation}
  Here $\lVert \cdot \rVert_F$ denotes the Frobenius norm,
  $\lvert \cdot \rvert_1$ denotes the entry-wise $1$-norm,
  and $\mu_W$, $\mu_H > 0$ are parameters controlling the
  sparsity of the matrices $W_i$ and $H_i$, respectively.
  We note that the training of a basis for NMF is a non-convex
  problem that usually admits multiple local solutions.
  This lack of convexity is a typical problem in the fitting of generative functions.
  \medskip

  In the case where we only have samples from the sources $1,\ldots,S-1$,
  but not from source $S$, the training is slightly different:
  For the training of the bases $W_i$, $i=1,\ldots,S-1$, of the given
  sources we proceed as above.
  Then, however, we estimate a basis $W_S$ for the last source
  by trying to fit the given samples to mixed data.
  To that end, we store the mixed data column\-wise in a matrix $V$
  and compute $W_S$ by solving
  \begin{equation}
    \min_{\substack{W_S \ge 0,\\H_i\ge 0,\, i=1,\ldots,S}}
    \Bigl\lVert V - W_S H_S - \sum_{i=1}^{S-1} W_i H_i\Bigr\lVert_F^2
    + \mu_W \lvert W_S\rvert_1 + \mu_H \sum_{i=1}^S \lvert H_i\rvert_1.
    \label{eq:semi}
  \end{equation}
  How well $W_S$ is able to approximate the unknown source $S$ then depends on the
  quality of the other bases $W_i$, $i=1,\ldots,S-1$, as well as the amount
  of available mixed data.  
\subsubsection*{Separation of new data}
  Given a mixed signal $v \in \mathbb{R}^m$ that we want to separate,
  we can solve the problem
  \begin{equation}
    (h_i^\ast)_{i=1}^S = \argmin_{h_i \ge 0,\, i = 1,\ldots,S}
    \Bigl\lVert V - \sum_{i = 1}^S W_i h_i\Bigr\rVert_F^2 + \mu_H \sum_i \lvert h_i \rvert_1
    \label{eq:gentest}
  \end{equation}
  and recover the separated signals as $u_i^\ast = W_i h_i^\ast$.
  For solving this problem, we can concatenate the bases and the latent
  variables to matrices
  \begin{equation*}
    W = \begin{bmatrix}W_1 & \hdots & W_S \end{bmatrix},
    \quad
    h = \begin{bmatrix}h_1^T & \hdots & h_S^T \end{bmatrix}^T,
  \end{equation*}
  and then solve the problem
  \begin{equation*}
    h^\ast = \argmin_{h \ge 0} \lVert V - Wh\rVert_F^2 + \mu_H \lvert h\rvert_1, 
  \end{equation*}
  which is a non-negative least squares problem with sparsity constraints,
  for which many efficient solution methods exist.
  
  In order to ensure that the recovered sources still sum to the given signal $v$,
  it is common to apply afterwards a Wiener-type filter
  \begin{equation*}
    u_i = v \odot \frac{W_i h_i^\ast}{\sum_{j = 1}^S W_jh_j^\ast},
  \end{equation*}
  where $\odot$ denotes entry\-wise multiplication the division is interpreted entry\-wise.

Historically, this approach has been named ``supervised NMF''
to distinguish it from the unsupervised/blind problem
where we do not have data from the individual sources.
In order to emphasize the difference to the strong supervision
case where we have access to mixed data together with
the correct separated sources, we denote this as
``standard NMF.''

One advantage of this approach is that we can decouple the training from
the SCSS problem we want to solve, as we can train the bases $W_i$ without 
any information about the joint distribution $\mathbb{P}_{\tilde{U} \times V}$.
Moreover, the testing problem is convex and therefore
not sensitive to initializations.
Also, the 1-homogeneity of the generator function
implies it can be used irrespective of the weights of the mixed signal we want to separate,
which cannot usually be said for more complex generative functions.

\subsection{Variants of NMF}

In this part, we will briefly introduce two variants
of the NMF method discussed above,
namely discriminative NMF, which is intended
for usage with strong supervision data,
and exemplar based NMF, where the basis vectors
are chosen as random samples of the data.

\subsubsection*{Discriminative NMF}

The approach discussed in Section~\ref{ss:NMFSS}
can also be extended to the case where strong supervision data
in the form of mixed data together with the correct separation
\[
\tilde{V} = \sum_{i=1}^S \tilde{U}_i
\]
is available.
Here, the data has already been collected column\-wise into matrices.
Given this data, we then solve the multi-objective optimization problem
\begin{equation}
  \min_{W \ge 0} \bigl(\lVert\tilde{U}_1 -  W_1 H_1(V,W)\rVert_F^2 ,\ldots,\lVert\tilde{U}_S - W_S H_S(V,W)\rVert_F^2\bigr) \label{eq:DNMFMO}
\end{equation}
where
\begin{equation*}
\bigl(H_1(\tilde{V},W),\ldots,H_S(\tilde{V},W)\bigr) =  
\argmin_{H_i \ge 0,\, i = 1,\ldots,S} \Bigl\lVert\tilde{V} - \sum_{i = 1}^S W_iH_i\Bigr\rVert_F.
\end{equation*}
This model, as well as the variants discussed in the following,
is called \emph{Discriminative NMF} (DNMF).

Usually, the problem~\eqref{eq:DNMFMO} is not solved in this
multi-objective form, but rather in a scalarized version
proposed in \cite{weninger2014discriminative,6843241}.
This scalarized version of~\eqref{eq:DNMFMO} can be written as the bi-level problem
\begin{equation}\label{eq:DNMF}
  \min_{W_i \ge 0,\, i = 1,\ldots,S}
  \sum_{i = 1}^S \gamma_i \bigl\lVert\tilde{U}_i -  W_i H_i(\tilde{V},W)\bigr\rVert_F^2
\end{equation}
with
\begin{equation*}
 \bigl(H_1(\tilde{V},W),\ldots,H_S(\tilde{V},W)\bigr) =  
 \argmin_{H_i \ge 0,\, i = 1,\ldots,S} \Bigl\lVert\tilde{V} - \sum_{i = 1}^S W_iH_i\Bigr\rVert_F,
\end{equation*}
where $\gamma_i > 0$, $i=1,\ldots,S$, are tuning parameters.
These need to be chosen a priori and assign weights to the sources depending
on which sources we are most concerned with reconstructing.
In \cite{weninger2014discriminative}, the authors solve a version of this problem
where $H_i(\tilde{V},W)$ is instead
chosen by solving \eqref{eq:NMFtrainingH} for the individual sources
and is held constant while solving \eqref{eq:DNMF}.
Thus, the problem is not longer bi-level. This means that the values of 
$\gamma$ are irrelevant, as the different terms of the sum become independent of each other. 
They also propose to change the upper level problem to directly optimize the Wiener-filtered solution, in which case the different
terms again do depend on each other.
 
Solving \eqref{eq:DNMF} is harder than training the bases $W_i$ individually, as we need to optimize over all bases at the same time.
However, we will propose a simple algorithm for solving this problem in Section~\ref{sec:numerical}, which is almost identical 
to fitting the bases individually with \eqref{eq:NMFtrainingW}.
This is in contrast to more advanced generative models where discriminative training can be infeasible
due to large search spaces.

\subsubsection*{Exemplar based NMF}

The main disadvantage of NMF is that it is unable to approximate non-stationary data and data with highly non-linear structures. 
Some authors have proposed adaptive sparsity and using convolutional dictionaries to deal with these difficulties \cite{gao2011adaptive}.
Another way of solving the problem is to train NMF with many basis vectors to be able to represent the data.
However, this again leads to the bases being able to represent signals of other sources, which in turn leads to poor separation results.
In the literature, there are mainly two ways of alleviating this problem: Using sparsity during testing and training, so that a signal can be represented with few basis vectors
\cite{le2015sparse}, and exemplar based NMF (ENMF), where the basis vectors are simply chosen as randomly sampled signals from the training set \cite{weninger2014discriminative}.
ENMF has the benefit that it requires no training (only sampling), but can achieve comparable results to standard NMF in some cases.

\section{Adversarial regularization for Single Channel Source Separation}
\label{sec:adversarial}

\subsection{Adversarial regularization for inverse problems}

In recent years, unsupervised and semi-supervised methods for training
regularization functionals for the inverse problem of solving an equation
\[
  Au = v
\]
have received much attention.
One prominent example is the article \cite{lunz2018adversarial}.
There, the authors assume that one has knowledge about the probability
distributions $\mathbb{P}_V$ and $\mathbb{P}_U$ of the measured data and the data of interest respectively.
In order to train a regularization functional, they use the probability
distribution $\mathbb{P}_V$ in order to define a new distribution of adversarial data on
the solution space by setting
\[
  \mathbb{P}_Z := (A^\dagger)_{\#} \mathbb{P}_U.
\]
Here $A^\dagger$ is the pseudo-inverse of the matrix $A$,
and $(A^\dagger)_{\#}$ denotes the push-forward operation.

Then they define a regularization functional $R$ for the solution
of the inverse problem by computing the \emph{Wasserstein distance}
$\mathbb{W}(\mathbb{P}_U,\mathbb{P}_Z)$ between the distributions
$\mathbb{P}_U$ and $\mathbb{P}_Z$.
More precisely, they set $R$ to be the solution of
\begin{equation}
  \mathbb{W}(\mathbb{P}_U, \mathbb{P}_Z) =
  \min_{R:\lVert R\rVert_L \le 1} \mathbb{E}_{u \sim \mathbb{P}_U}[R(u)] - \mathbb{E}_{u \sim \mathbb{P}_Z}[R(u)],
  \label{eq:advregtrain}
\end{equation}
where $\lVert R\rVert_L$ denotes the Lipschitz constant of $R$
and $\mathbb{E}_{u\sim \mathbb{P}_X}$ the expectation given the distribution $\mathbb{P}_X$.
Although~\eqref{eq:advregtrain} would require the minimization
of the right-hand side over all functions $R$ with Lipschitz constant bounded by $1$,
in practice, one restricts $R$ to be of some a-priori chosen class,
for instance neural networks with a given architecture.
In order to obtain a convex regularization term $R$, it was further proposed
in~\cite{mukherjee2020learned} to restrict the neural networks to
ones with convex activation functions and positive weights.

The intuition behind the adversarial approach is that $R$ should yield low values for true solutions and
large values for naively inverted data, which can be treated as adversarial data, since
the measurement data $v$ can be assumed to include noise.
Importantly, though, it does not require information about the
joint distribution $\mathbb{P}_{U \times V}$, and fits into the weak supervision data setting.

After training the regularization function $R$, it can be used during testing by solving
the familiar regularization problem
\begin{equation}
 \min_{u} \lVert Au - v\rVert_2^2 + \lambda R(u),
 \label{eq:advregtest}
\end{equation}
where $\lambda > 0$ is a regularization parameter.
This adversarial approach is in contrast to the discriminative approach for solving inverse problems,
where the goal is to learn an operator from the measured signal $v$
or its pseudo-inverse $A^\dagger v$ to the unknown inverse $u$ \cite{jin2017deep}.
Adversarial regularization functions potentially lead to flexible regularization functions $R$, 
that require much fewer fitted parameters compared to directly
learning mappings between large spaces discriminatively.

\subsection{Application to source separation}

We will now propose a strategy for adapting the
idea of adversarial regularization to source separation
with weak supervision.
Although source separation can be regarded as an inverse problem,
there are some significant differences to the setting considered
in~\cite{lunz2018adversarial} that make these adaptations necessary.

First, we note that~\eqref{eq:advregtrain} requires
knowledge of the joint distribution $\mathbb{P}_U$ of
the different sources, which is in general not available in
a weak supervision setting.
Instead, we only have access to the marginal distributions $\mathbb{P}_{U_i}$
for each of the sources.
As a consequence, we propose to train a different regularization
term $R_i$ for each of the sources and to write $R$ as convex combination
\begin{equation*}
  R(u_1,\ldots,u_S) = \sum_{i=1}^S \rho_i R_i(u_i)
\end{equation*}
with weights $\rho_i > 0$ satisfying $\sum_{i=1}^S \rho_i = 1$.

Moreover, we propose to train the terms separately.
That is, we define the regularization term $R_i$ as
\[
  R_i = \argmin \mathbb{E}_{u \sim \mathbb{P}_{U_i}}[R_i(u)]
  - \mathbb{E}_{u \sim \mathbb{P}_{Z_i}}[R_i(u)],
\]
where we take the minimum over a suitable class of functionals,
which we will specify in the next section.
Here $\mathbb{P}_{Z_i}$ denotes the probability distribution
of the adversarial data that are used for the training
of the $i$-th regularization term.

Regarding the choice of the adversarial data,
we follow the argumentation of~\cite{lunz2018adversarial}
and include data that is produced by naively inverting
the forward operator $A$, as this approach is expected
to preserve, or even amplify, possible noise in the data.
For this, we need the pseudo-inverse of $A$, which,
for the case of source separation, becomes
\[
  A^\dagger v = \Bigl(\frac{a_1}{\sum_{j = 1}^S a_j^2} v,\ldots,
  \frac{a_S}{\sum_{j = 1}^S a_j^2} v\Bigr).
\]
This operator, however, depends on the weights $0 \le a_j \le 1$,
which we assume to be randomly distributed according
to some distribution $\mathbb{P}_A$ on the space of weights.
We now define
\[
  f_i(a_1,\ldots,a_S;v) = \frac{a_i}{\sum_{j = 1}^S a_j^2} v,
\]
which is precisely the $i$-th component of $A^\dagger v$.
Then we obtain the distribution $\mathbb{P}_{V_i}$ of
the $i$-th component of naively inverted mixed data
as the push-forward of the joint distribution $\mathbb{P}_{A\times V}$
of weights and mixed data via the mapping $f_i$, that is,
\[
  \mathbb{P}_{V_i} = (f_i)_\# (\mathbb{P}_{A\times V}).
\]
In the weak supervision setting, we cannot generally assume
that the joint distribution $\mathbb{P}_{A\times V}$ is available.
Lacking any better model, we will therefore assume that
$\mathbb{P}_A$ and $\mathbb{P}_V$ are independent.
Moreover, for the distribution of the weights, a natural choice of the distribution
is the Dirichlet distribution. Alternatively, we can simply assume they are deterministic.

In addition to naively inverted mixed data, we propose
to use data from other sources $j \neq i$ for the training of $R_i$,
as we intuitively want $R_i(u)$ to yield large values for these
type of data.
Thus, we sample the adversarial data from the mixture distribution
\[
  \mathbb{P}_{Z_i} := \sum_{j\neq i} \omega_{ij} \mathbb{P}_{{U}_j}
  + \Bigl(1-\sum_{j\neq i} \omega_{ij}\Bigr)\mathbb{P}_{V_i},
\]
where the parameters $\omega_{ij}$ determine the weight
of the different adversarial sources.
A natural choice for these weights are the ratios
\begin{equation*}
  \omega_{ij} = \frac{N_j}{\hat{N}_i},
  \qquad\text{ where } \hat{N}_i := N_V + \sum_{k \neq i} N_k
\end{equation*}
denotes the total amount of adversarial data for source $i$.
Other choices of $\omega_{ij}$ can be interesting, though,
especially in the case where the different sources contain unbalanced data.

\subsection{Adversarial Generative NMF}
\label{sec:ANMF}

We will now apply the idea of adversarial regularization
to non-negative matrix factorization.
Here we want to model data from the source $i$ by a generating function
$g_i\colon \mathbb{R}_+^d \to \mathbb{R}_+^m$ of the form
$g_i(h) = W_i h$ where $W_i \in \mathbb{R}_+^{m\times d}$
is the collection of (positive) basis vectors for that source.
With a generating function, we can solve the test
problems~\eqref{eq:advregtest} and recover a solution $u^\ast$ as 
\begin{equation}
  h^\ast = \argmin_{h \in \mathcal{H}} \lVert Ag(h) - v\rVert^2, \quad u^\ast = g(h^\ast),
\end{equation}
which has the same form as the test problem for
source separation \eqref{eq:gentest}.
Using this approach for inverse problems is not novel \cite{duff2021regularising},
but the coupling between generative models and adversarial regularization
functions is novel.
We call this approach \emph{Adversarial Generative Regularization},
or, more specifically for the case where $g_i(h) = W_i h$,
\emph{Adversarial Generative Non-Negative Matrix Factorization} (ANMF).

We note here that the name \emph{Adversarial Non-Negative Matrix Factorization} has already been proposed by other authors \cite{luo2020adversarial},
but for a different problem setting:
In \cite{luo2020adversarial}, the main concern is fitting
NMF to be robust against adversarial attacks, that is,
robust against small perturbations in the data
that can lead to vastly different reconstructions.
In contrast to this, we are in this article interesting in training NMF
that fits selected adversarial data poorly, to be used in solving
source separation and inverse problems.

Denote now by $C(W_i) = g_i(\mathbb{R}_+^d)$ the convex cone
spanned by the columns of $W_i$.
We propose to parameterize the regularization functions $R_i$ as
\[
  R_i(u) = \lVert u - P_{C(W_i)}(u) \rVert = D_{C(W_i)}(u),
\]
where $P_{C(W_i)} \colon \mathbb{R}^m \to \mathbb{R}^m$
is the projection onto the convex cone $C(W_i)$
and $D_{C(W_i)}$ the distance to $C(W_i)$.
This function has many desirable properties, such as: convexity, $1$-Lipschitz continuity, positive $1$-homogeneity, and it is differentiable everywhere
except on $\partial C(W_i)$.

Instead of plugging this parametization directly into the training problem for adversarial regularization functions \eqref{eq:advregtrain},
we instead propose minimizing the squared distances
\begin{equation}
\min_{W_i \ge 0} \mathbb{E}_{u \sim \mathbb{P}_{U_i}}[D_{C(W_i)}(u)^2] - \mathbb{E}_{u \sim \mathbb{P}_{Z_i}}[D_{C(W_i)}(u)^2],
\label{eq:squared}
\end{equation}
as this problem is differentiable everywhere, and the gradients are simpler.
If we ignore the second term of \eqref{eq:squared},
this reduces to the standard NMF formulation when we apply 
Monte Carlo integration for evaluating the expected value.
The main difference between ANMF and NMF is that we are
not only concerned with fitting data well, but we are also
interested in fitting adversarial data poorly.  

For the calculation of the second expectation in~\eqref{eq:squared},
it is necessary to take samples from the mixed data
and the distribution $\mathbb{P}_A$ of weights.
However, due to the $1$-homogeneity of $D_{C(W_i)}$ and the assumed independence
of $\mathbb{P}_V$ and $\mathbb{P}_A$, this term simplifies to
\begin{multline*}
  \mathbb{E}_{u \sim \mathbb{P}_{Z_i}}[D_{C(W_i)}(u)^2] \\
    =\sum_{j \neq i}\mathbb{E}_{u \sim \mathbb{P}_{U_j}}[D_{C(W_i)}(\sqrt{\omega_{ij}}u)^2] 
     + \mathbb{E}_{u \sim \mathbb{P}_V}
    \Bigl[D_{C(W_i)}\Bigl(\Bigl(1 - \sum_{j \neq i}\omega_{ij}\Bigr)^{1/2}\beta_i^{1/2} u\Bigr)^2\Bigr]
\end{multline*}
with
\begin{equation}
  \beta_i = \mathbb{E}_{a \sim \mathbb{P}_A}\Bigl[\Bigl(a_i / \Bigl(\sum_{j} a_j^2\Bigr)\Bigr)^2\Bigr].
  \label{eq:beta}
\end{equation}

In order to stabilize the method, and also the
subsequent numerical algorithms, we add
additional regularization terms both to the
upper and the lower level problems.
With this addition, we then can write the training problem using Monte Carlo integration
conveniently in matrix form as
\begin{equation}\label{eq:ANMF}
  \min_{W_i \ge 0} \frac{1}{N_i} \lVert U_i - W_iH(U_i, W_i)\rVert_F^2 
  - \frac{1}{\hat{N}_i} \lVert \hat{U}_i - W_iH(\hat{U}_i, W_i)\rVert_F^2 + \mu_W \lvert W_i\rvert_1 ,
\end{equation}
where
\begin{equation*}
H(U,W) = \argmin_{H\ge 0} \lVert U - WH\rVert_F + \mu_H \lvert H\rvert_1.
\end{equation*}
Here $\hat{U}_i$ is the adversarial data of the $i$-th
source, stored column\-wise and scaled as
\begin{equation*}
\hat{U}_i =  \begin{bmatrix}\alpha_1 U_1 & \hdots & \alpha_{i-1} U_{i-1}& \alpha_{i+1} U_{i+1} & \hdots & \alpha_S U_S & \alpha_V V 
\end{bmatrix},
\end{equation*}
where 
\begin{equation}\label{eq:alphaj}
\alpha_j =\sqrt{\frac{\omega_{ij}\hat{N}_j}{N_j}}, \quad \alpha_V =\sqrt{\frac{\bigl(1 - \sum_{j \neq i}\omega_{ij}\bigr) \hat{N}_j \beta_i}{N_V}}. 
\end{equation}
In the case where $\omega_{ij} = N_j/\hat{N}_i$, this is just the concatenation of all the data from other sources $U_j$, $j \neq i$, and naively inverted mixed data $\sqrt{\beta_i} V$.

We note here that, similarly as for standard NMF,
the training problem for ANMF in~\eqref{eq:ANMF} is non-convex.
Moreover, it is invariant under rescalings of the columns of $W_i$,
as these can simply be absorbed by corresponding rescalings of
the rows of $H$.
It therefore makes sense to include the additional constraint that the
columns of $W$ should be scaled to norm $1$.

\subsubsection*{Weighted ANMF}
There is a clear trade-off between goodness of fit of the true data and poorness of fit of the adversarial data, and it is possible that the adversarial term dominates
the optimization.
In the worst case, we can end up with a (local or even global)
minimizer $W_i$ for which $W_i H_i = 0$.
It is also clear that the ANMF basis will always do a worse job at representing training data than standard NMF.
For low complexity methods like NMF, this means true data will be fitted poorly to the point that the bases will be unusable for source separation or inverse problems
when the adversarial term dominates.

To alleviate the problem of the adversarial term dominating, as well as to stabilize the problem, we modify the problem and fit a weighted mix between standard NMF and ANMF,
\begin{multline}
  \min_{W_i \ge 0} (1- \tau_A )\Bigl(\mathbb{E}_{u \sim \mathbb{P}_{U_i}}[D_{C(W_i)}(u)^2] + \mu_{W} \lvert W_i \rvert\Bigr)\\
  + \tau_A \Bigl(\mathbb{E}_{u \sim \mathbb{P}_{U_i}}[D_{C(W_i)}(u)^2]
  - \mathbb{E}_{u \sim \mathbb{P}_{Z_i}}[D_{C(W_i)}(u)^2] + \mu_{W} \lvert W_i \rvert\Bigr)\\
  = \min_{W_i \ge 0} \mathbb{E}_{u \sim \mathbb{P}_{U_i}}[D_{C(W_i)}(u)^2]
  - \tau_A \mathbb{E}_{u \sim \mathbb{P}_{Z_i}}[D_{C(W_i)}(u)^2] + \mu_{W} \lvert W_i \rvert,
\label{eq:WANMF}
\end{multline}
where $\tau_A \ge 0$ is a tuning parameter that must be chosen a priori. The role of $\tau_A$ is selecting how concerned we are with fitting adverserial data poorly,
with small $\tau_A$ leading to better fits for true data and larger $\tau_A$ leading to worse fits for adversarial data. We therefore call $\tau_A$ the \textit{adversarial weight}, and it can potentially be chosen differently for each source.
Selecting $\tau_A = 0$ means we are just fitting a standard NMF.

The parameter $\tau_A$ must be chosen a priori with some heuristic or tuned with hyperparameter tuning methods.
In this article we will mainly be interested in selecting it using the latter approach, as well as investigating the impact of $\tau_A$.

\subsection{Discriminative and Adversarial Generative NMF}
\label{sec:FNMF}

We can easily extend the ANMF model to also take into account strong supervised data
by adding an extra term corresponding to DNMF.
Assume to that end that we are given weak supervised data in the
form of probability distributions $\mathbb{P}_{U_i}$, $i = 1,\ldots,S$,
as well as strong supervised data in the form of a joint probability
distribution $\mathbb{P}_{U\times V}$.
Moreover, choose weights $0 \le \tau_A,\, \tau_S \le 1$
for the adversarial term and the strong supervised term, respectively.
Then we can define for $i = 1,\ldots,S$ the functional
\begin{multline*}
  F_i(W) = (1-\tau_S) \underbrace{\mathbb{E}_{u_i \sim \mathbb{P}_{U_i}}\bigl[D_{C(W_i)}(u_i)^2\bigr]}_{\text{Weak supervised}}
        + (1 - \tau_S)\tau_A \underbrace{\mathbb{E}_{u_i \sim \mathbb{P}_{Z_i}}\bigl[D_{C(W_i)}(u_i)^2\bigr]}_{\text{Adversarial}}\\
  +\tau_S \underbrace{\mathbb{E}_{(u_1,\ldots,u_s,v) \sim \mathbb{P}_{U \times V}}
            \Bigl[\bigl\lVert u_i - W_ih_i(v,W)\bigr\rVert_F^2\Bigr]}_{\text{Strong supervised}},
\end{multline*}
where, again, $\mathbb{P}_{Z_i}$ denotes the distribution of the adversarial
data for the $i$-th source, and
\[
\bigl(h_1(v,W),\ldots,h_S(v,W)\bigr) = \argmin_{h_i \ge 0,\, i = 1,\ldots,S} \Bigl\lVert v - \sum_{i = 1}^S W_ih_i\Bigr\rVert.
\]
We then obtain the multi-objective optimization problem
\begin{equation*}
  \min_{W_i \ge 0,\, i=1,\ldots,S} \bigl(F_1(W),\ldots,F_S(W)\bigr),
\end{equation*}
which, as for the case of DNMF, can also be rewritten in
a weighted sum formulation as
\begin{equation}\label{eq:FNMF}
  \min_{W_i \ge 0,\, i=1,\ldots,S} \sum_{i=1}^S \gamma_i F_i(W).
\end{equation}
This leads to a general framework for fitting NMF that is flexible
with respect to all available data. We call this approach \emph{Discriminative and Adversarial Generative NMF} (D+ANMF), and it encompasses the NMF models discussed so far,
which can be recovered by setting certain parameters to $0$ or $1$,
see Table \ref{tab:NMF}.
\begin{table}[!hbt]
\centering
\begin{tabular}{c|cccc}
& NMF & ANMF & DNMF & D+ANMF \\ 
\hline
$\tau_A$ & $=0$ & $>0$ & $=0$ & $\ge 0$ \\ 
$\tau_S$ & $=0$ & $=0$ & $=1$ & $\in(0,1)$
\end{tabular}
\caption{An illustration of the parameter values of $\tau_A$ and $\tau_S$ for the different variations of NMF. 
D+ANMF is a superset of all methods. }
\label{tab:NMF}
\end{table}

This approach is most interesting in the case where only a small amount
of strong supervised data is available as compared to weak supervised data.
In situations with large amounts of strong supervised data or suitable
synthetic supervised data, one can expect that fitting purely discriminatively
or purely adversarially is a better approach.
We also have the option of using the strong supervised data for fitting
both the weak supervision terms and strong supervision terms. For D+ANMF, this would mean that $\tau_S$ represents the relative weight of weak supervision compared to
strong supervision fitting. This can potentially alleviate the problem of overfitting to specific data.
Note also, that strong supervised data can alternatively be used
for hyperparameter fitting. If only a small amount of strong supervised
data is available, it can then be advantageous, not to use it for strong supervised fitting,
but for hyperparameter tuning of weak supervision models.

\section{Numerical implementation}
\label{sec:numerics}

We now propose a numerical multiplicative algorithm for fitting ANMF and D+ANMF similar to a standard algorithm for fitting NMF proposed in \cite{lee2000algorithms}.

\subsection{Multiplicative updates}

The standard approach to fitting NMF is to alternatively update the basis while keeping the latent variables fixed, 
and then update the latent variables while keeping the basis fixed. This approach can usually find a local minimizer.

A standard multiplicative update for finding $H^\ast$ so that $P_{C(W)}(U) = WH^\ast$ is given by 
\begin{equation}
H \leftarrow H \odot \frac{W^TU}{W^TWH + \mu_H},
\label{eq:Hupdate}
\end{equation}
where $\mu_H$ is both a safe division factor and a sparsity parameter for $H$ \cite{lee2000algorithms}.
Here entry\-wise multiplication (the Hadamard product) is denoted by $\odot$,
and the (Hadamard) division is interpreted entry\-wise as well. 
We can obtain a similar update rule for $W$, which will be introduced later, and solve the NMF training problem \eqref{eq:NMFtrainingW}
by alternatively updating $H$ and $W$.

\subsubsection*{Multiplicative update for ANMF}

We denote here by $H_i$ the latent variables for the weak supervision term of the $i$-th source, and by $\hat{H}_i$ the latent variables of the adversarial term of the $i$-th source.
We use a similar notation for
the data: $U_i \approx W_i H_i$ and $\hat{U}_i \not\approx W_i \hat{H}_i$. 
The update can be written by first splitting the gradient into positive and negative terms,
then applying a multiplicative update as follows:
\begin{align}
[\nabla W_i]_\text{std}^+ &= W_iH_iH_i^T/N_i, &
[\nabla W_i]_\text{std}^- &= U_iH_i^T/N_i, \label{eq:ANMFWstart} \\
[\nabla W_i]_\text{adv}^+ &= \tau_A \hat{U}_i\hat{H}_i^T/\hat{N}_i, &
[\nabla W_i]_\text{adv}^- &= \tau_A W_i\hat{H}_i\hat{H}_i^T/\hat{N}_i,\\
W_i \leftarrow W_i &\odot \frac{[\nabla W_i]_\text{std}^- + [\nabla W_i]_\text{adv}^-}{[\nabla W_i]_\text{std}^+ + [\nabla W_i]_\text{adv}^+ + \mu_W}.
\label{eq:ANMFWend}
\end{align}
Similarily, the updates for the latent variables are given by applying~\eqref{eq:Hupdate} to the appropriate data:
\begin{equation}
  H_i \leftarrow H_i \odot \frac{W_i^TU_i/N_i}{W_i^TW_iH_i/N_i + \mu_H},
  \qquad\qquad
  \hat{H}_i \leftarrow \hat{H}_i \odot \frac{W_i^T\hat{U}_i/\hat{N}_i}{W_i^TW_i\hat{H}_i/\hat{N}_i + \mu_H}. \label{eq:FNMFHadv}
\end{equation}

In order to deal with the scale invariance of the problem (see Section~\ref{sec:ANMF}),
we normalize the columns of $W$ between the epochs.
That is, after each epoch, we divide each column of $W$
by its norm and simultaneously multiply each row of the latent variables
by the same number.
Thus, the products $WH$ and $W\hat{H}$ are
unchanged, but the columns of $W$ are normalized.

\subsubsection*{Multiplicative update for D+ANMF}
We can extend the update to also account for a discriminative term.
To that end, we denote by $\tilde{H_i}$ the latent variables of the data
for the strong supervision term of the $i$-th source,
and similarly by $\tilde{V} = \sum_{i = 1}^S\tilde{U}_i \approx \sum_{i = 1}^S W_i \tilde{H}_i = W\tilde{H}$ the supervised data.
The positive and negative parts of the gradient then become
\begin{equation}
  [\nabla W_i]_\text{sup}^+ = W_i\tilde{H}_i\tilde{H}_i^T/N_\text{sup},
  \qquad\qquad
  [\nabla W_i]_\text{sup}^- = \tilde{U}_i \tilde{H}_i^T/N_\text{sup},
  \label{eq:DNMFend}
\end{equation}
and the update for the concatenated latent variable $\tilde{H}$ is
\begin{equation}
  \tilde{H} \leftarrow \tilde{H} \odot \frac{W^T\tilde{V}/N_\text{sup}}{W^TW\tilde{H}/N_\text{sup} + \mu_H}, 
  \label{eq:FNMFHsup}
\end{equation}
where $W$ is the matrix obtained by concatenating the bases.

We can combine \eqref{eq:DNMFend} with equations \eqref{eq:ANMFWstart}--\eqref{eq:ANMFWend} to obtain the update for D+ANMF
\begin{equation}
W_i \leftarrow W_i \odot \frac{(1-\tau_S) [\nabla W_i]_\text{ANMF}^-+ \tau_S[\nabla W_i]_\text{sup}^-}{(1-\tau_S)[\nabla W_i]_\text{ANMF}^+ + \tau_S[\nabla W_i]_\text{sup}^+ + \mu_W},
\label{eq:fullWupdate}
\end{equation}
where we have used the abbreviations
\[
    [\nabla W_i]_{\text{ANMF}}^+ = [\nabla W_i]_{\text{std}}^+ + [\nabla W_i]_{\text{adv}}^+,\qquad\qquad
    [\nabla W_i]_{\text{ANMF}}^+ = [\nabla W_i]_{\text{std}}^+ + [\nabla W_i]_{\text{adv}}^-.
\]

The asymptotic computational complexity of each iteration of D+ANMF for all bases
is of order $\mathcal{O}(dmN_\text{tot})$, where $N_\text{tot} = \sum_{i = 1}^S [N_i +\hat{N}_i + N_\text{sup}]$ is the total amount of data.
Thus the computational complexity of the updates for NMF, ANMF, DNMF, and D+ANMF scales at the same rate with the amount of data. 

\subsubsection*{Initialization}

Because of the non-convexity of the problems we are trying to solve, the proposed updates
can only be expected to converge to a local minimizer.
Thus, a suitable initialization is required in order to obtain a good solution.
We will primarily use exemplar-based initialization, and randomized initialization
when exemplar-based initialization is not feasible.

\subsubsection*{Semi-supervised update}

In the semi-supervised case where the $S$-th source is unknown, but we have access to the mixed data $V$, the updates for $W_S$ and the latent variables $H_i$ become
\begin{equation*}
  W_S \leftarrow W_S \odot \frac{V H_S^T/N_V}{(\sum_{i = 1}^S W_i H_i) H_S^T/N_V + \mu_W}, \quad
  H_i \leftarrow H_i \odot \frac{W_i^T V/N_V}{W_i^T(\sum_{j = 1}^S W_j H_j)/N_V + \mu_H}.
\end{equation*}
Here the bases $W_i$, $i = 1,\ldots,S-1$, are pre-trained, and they can be trained adversarially by using the mixed data as adversarial data.

\subsection{Stochastic Multiplicative Updates}
\label{sec:multiplicative_update}

For standard NMF, each column of the data $U$ has a corresponding column in the latent variable $H$. At the start of each epoch,
we can shuffle the data in $U$, perform a corresponding shuffle of $H$, and divide the matrices column\-wise into batches $U^{(b)}$ and $H^{(b)}$ following the ideas of \cite{serizel2016mini}.
We can then successively apply the update for $W$ for data from the different batches.
We update all latent variables $H$ simultaneously instead of batch-wise in a single update.
To stabilize the ANMF algorithm, we found it helpful to normalize $W$ before updating $H$. 
We call this method Stochastic Multiplicative Update (SMU).

When applying SMU to ANMF and D+ANMF we face the challenge that we are minimizing a loss with
different terms and potentially unbalanced data, and need to select batch sizes accordingly. 
To overcome this, we select one term we are interested in fully sampling and undersample or oversample data from the other terms.
The data is shuffled and a new epoch begins when we have passed through all data of this chosen term.
The full proposed algorithm for fitting D+ANMF where the same parameters are used for all sources is given in Algorithm~\ref{alg:smu}.

\begin{algorithm}
\caption{Stochastic Multiplicative Update for D+ANMF}
\label{alg:smu}
\begin{algorithmic}
  \STATE \textbf{Input: } $\text{epochs} \in \mathbb{N}$, $d \in \mathbb{N}$, $\mu_H$, $\mu_W > 0$, $\tau_A$, $\tau_S > 0$, and batch sizes.
  \STATE \textbf{Data input:} True, adversarial, and supervised datasets $U \in \mathbb{R}_+^{S \times m \times N}$, $\hat{U} \in \mathbb{R}_+^{S \times m \times \hat{N}}$, and $\tilde{U} \in \mathbb{R}_+^{S \times m \times N_{\text{sup}}}$. Supervised mixed data $\tilde{V} \in \mathbb{R}_+^{m \times N_\text{sup}}$.
  \STATE \textbf{Initialize: } $W \in \mathbb{R}_+^{m \times d}$ randomly or exemplar-based.
  \STATE \textbf{Initialize: } Latent variables $H$, $\hat{H}$, $\tilde{H}$ either randomly or with $\eqref{eq:Hupdate}$ applied to the respective data.
  \STATE \textbf{Calculate: } Number of batches.
  \FOR{$k = 0, k < \text{epochs}$}
    \STATE Shuffle columns of $U, \hat{U}, \tilde{U}, H, \tilde{H}, \hat{H}, \tilde{V}$.
    \STATE Update $\tilde{H}$ with \eqref{eq:FNMFHsup}.
    \FOR{$i = 1, i \le S$}
      \STATE Update $H_i$ and $\hat{H}_i$ with \eqref{eq:FNMFHadv}. 
      \FOR{$b = 0, b < \text{number of batches}$}
        \STATE Update $W_i$ with \eqref{eq:fullWupdate} using $U_i^{(b)}$, $H_i^{(b)}$, $\hat{U}_i^{(b)}$, $\hat{H}_i^{(b)}$, $\tilde{U}_i^{(b)}$ and $\tilde{H}_i^{(b)}$.
      \ENDFOR
    \ENDFOR
    \STATE Normalize $W$, $H$, $\hat{H}$, $\tilde{H}$.
  \ENDFOR
\end{algorithmic}
\end{algorithm}

This universal algorithm for D+ANMF can also be used to fit NMF, ANMF and D+ANMF by selecting the parameters $\tau_A$ and $\tau_S$, see Table~\ref{tab:NMF}. 
It is also worth noting that we can swap the order of the loop over sources and the loop over epochs. For NMF and ANMF this does not affect anything,
as the bases can be fitted independently of each other in parallel. This should however not be done for DNMF and D+ANMF, as the bases should be updated concurrently for each epoch.

One unfortunate property of ANMF and D+ANMF is that we need to update the latent variables for data that is not very relevant for the overall fit.
This is in particular the case if $\tau_A$ and $\tau_S$ are chosen low.
For practical applications, it would thus be possible to select only a subset
of the available data corresponding to the values of $\tau_A$ and $\tau_S$.
In this article, we still utilize and update all available data.

\subsection{Hyperparameter tuning}
\label{sec:hyperparameter}

Hyperparameter refers to a parameter that is chosen a priori and is not fitted during training. This usually needs
to be done in the presence of strong supervised data, even if the models themselves do not need to be trained with strong supervised data.
In this sense, performing hyperparameter tuning fits a discriminative model, because the model is tuned to a specific problem.

We will be interested in fitting several parameters at the same time, and we will do this where limited amounts of strong supervised data is available.
For this, we will apply a randomized search, as it is easy to implement while still being relatively efficient \cite{bergstra2012random}.
The idea of this method is to simply sample hyperparameters from predetermined probability distributions and select the parameters
that yield the best solution on the test data with or without cross-validation (CV) \cite{hastie01statisticallearning}.

For NMF and ANMF, which only require weak supervised data, we have the option of using the strong supervised
data for fitting or only for tuning. We assume that the benefits of using more data outweighs the detriments of overfitting,
and therefore use all data for fitting.

For DNMF and D+ANMF, we use CV to avoid overfitting the specific strong supervised data.
For D+ANMF, each CV fold use all weak supervision data available.

For source separation problems, the selection of a suitable metric
to be optimized during hyperparameter tuning
is a non-trivial problem, because we have to assess the
quality of all the separated sources simultaneously.
For that, we suggest using a weighted mean of some metric of interest
(PSNR, SDR, etc.) over the sources, where the weight is chosen
based on the importance of the sources.
When the signals are of equal importance, we
will use the arithmetic mean,
and for denoising problems we will ignore the noise part
and only weigh the signal of interest. 

\section{Numerical experiments}
\label{sec:numerical}

We now want to test our proposed algorithm for both image and audio data with a few different data settings.
For all experiments we implement the algorithms in Python using NumPy \cite{harris2020array}, and the code as well as supplementary material is available in the GitHub repository \url{https://github.com/martilud/ANMF}.

\subsection{Image data}
We first test our algorithms on the famous MNIST dataset \cite{deng2012mnist}. 

The MNIST dataset consists of $70000$ $28\times28$ grayscale images of $10$ different handwritten digits. 
We treat each of the different digits as a class, and attempt to separate mixed images
from a subset of the classes. We are only interested in the case where we know what classes a mix consists of.
Mixed data is generated independently, so technically 
we could use our training data to generate more mixes synthetically, but this is not something we will investigate.
See Figure~\ref{fig:data_rich_imgs} below for an example of a mixed image
together with the ground truth and reconstructed separated images.

For more information about the parameters used in the numerical experiments, as well as an experiment
that illustrates the convergence of our proposed numerical algorithm,
see the supplementary material.

\subsubsection*{Experiment 1: Data Rich, Strong Supervised Setting}

We investigate first a setting with an abundance of strong supervision data,
which make discriminative models like DNMF applicable.
In fact, we would expect DNMF to outperform the other methods
that do not make use of the fact that the data is strongly supervised.

We use $N_{\text{sup}} = 5000$ data-points and synthetically generate strong supervised data with ``zero'' and ``one'' digits with deterministic weights $a_0 = a_1 = 0.5$.
Similarly, we create $N_{\text{test}} = 1000$ data points that will be used for testing. 
We select the same number $d$ of basis vectors for both sources,
and test different values.

\begin{figure}[t!]
\centering
\includegraphics[width = 0.8\textwidth]{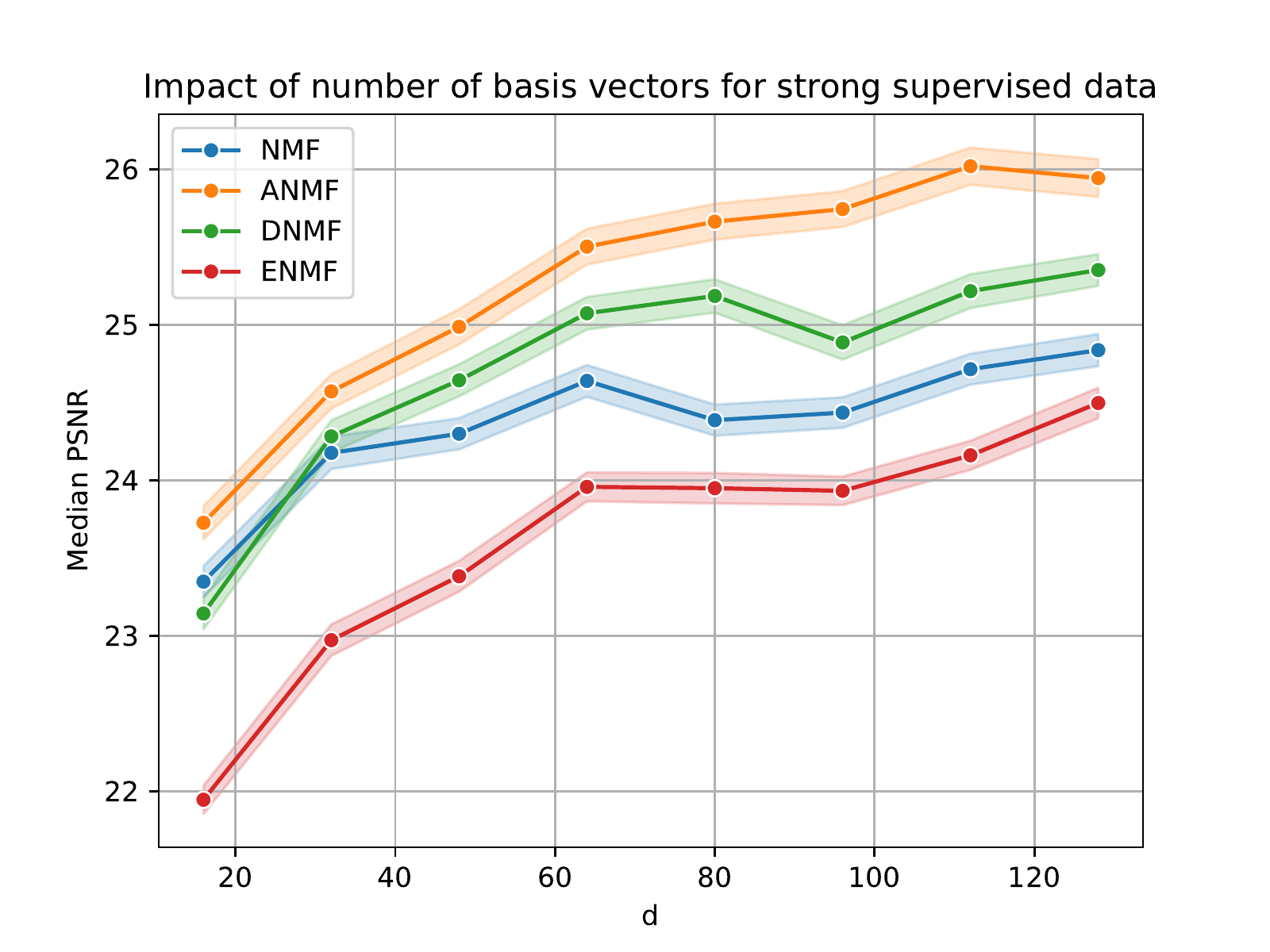}
\caption{Results from experiments in a data rich strong supervised setting. The lines show the median PSNR of the test dataset, along with the standard error.
For ANMF we use $\tau_A = 0.1$.
We note that performance tends to improve as the number $d$ of basis vectors increases, and that ANMF consistently outperforms the other methods.
For large $d$, the standard NMF bases become too flexible
and allow for the representation of different digits as well,
which leads to a decrease in performance. 
For ANMF, this does not happen, and performance instead increases
consistently with $d$, apart from a drop for $d = 128$.
We suspect that this performance drop in all methods expect ENMF
is caused by the training not properly converging.
DNMF behaves a bit erratically, which might be due to local minimizers.
The performance of ENMF also improves significantly as $d$ increases. }
\label{fig:data_rich}
\end{figure}

The results are illustrated in Figure~\ref{fig:data_rich}. We see that ANMF outperforms all methods, and
performance increases with the number of basis vectors $d$. 

Surprisingly, ANMF outperforms DNMF in this strong supervision setting,
even though it does not utilize the fact that we have strong supervision data, which DNMF explicitly does.
We predict that given enough data, DNMF can potentially find a global minimizer that outperforms ANMF, though this does not practically happen in our experiments.

We observe in the experiments that the performance of ENMF improves further as $d$ increases,
and it usually outperforms standard NMF for large $d$. This is remarkable, as ENMF requires
virtually no training.
However, since we are mainly interested in the case of lower values of $d$,
where ENMF performs poorly, we will omit the results of ENMF in the further experiments.

\begin{figure}[t!]
\centering
\includegraphics[width = 0.8\textwidth]{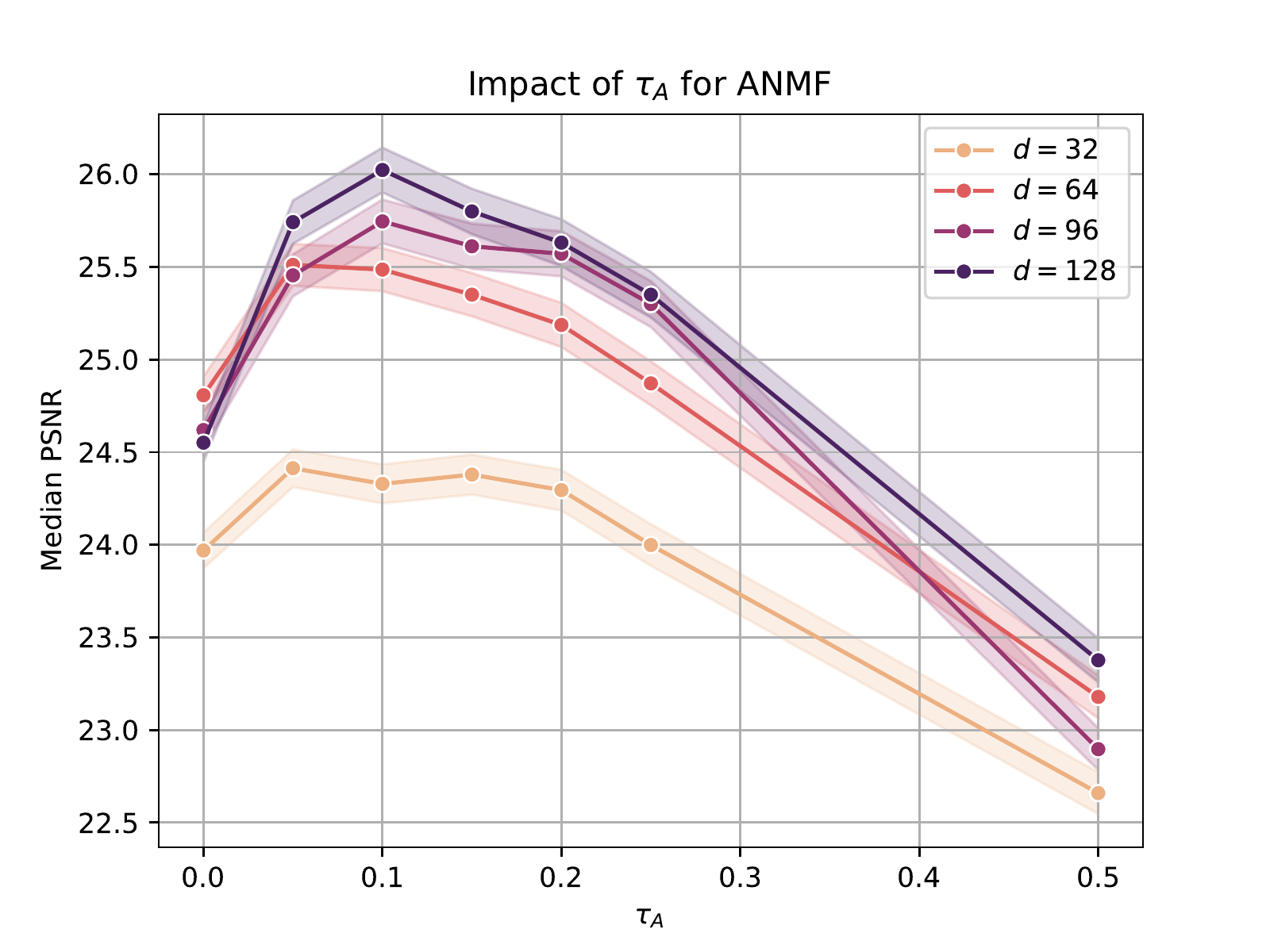}
\caption{Results from experiments in data rich strong supervised setting with varying $\tau_A$ for ANMF.
The lines show the median PSNR over the test dataset for different parameter values, along with the standard error.
We note that $\tau_A = 0$ corresponds with standard NMF. We see that selecting $\tau_A$ too large leads to much worse performance,
but there is a relatively large range of parameters that yield better performance than standard NMF. We also find that
the discrepancy in performance between NMF and ANMF becomes larger as $d$ increases.}
\label{fig:data_rich_tau}
\end{figure}

We now investigate how the performance is affected by the value of the parameter $\tau_A$. We run the same experiment,
except this time we vary $\tau_A$ and focus only on ANMF. The results are shown in Figure \ref{fig:data_rich_tau}.
They indicate that, while a good selection of the parameter $\tau_A$ is crucial for performance,
there is a relatively large range of values for which the performance is acceptable.
Noting that the case $\tau_A = 0$ corresponds to standard NMF,
we also see that the performance benefit of using ANMF over NMF increases with model complexity.

\begin{figure*}[t!]
\centering
\begin{tabular}{@{\hspace{-1mm}}c@{\hspace{-1mm}}c@{\hspace{-1mm}}c@{\hspace{-1mm}}c@{\hspace{-1mm}}c@{\hspace{-1mm}}}
  \textbf{Mix}
  & \textbf{Ground Truth} 
  & \textbf{NMF}  
  & \textbf{ANMF}
  & \textbf{DNMF}\\
  \includegraphics[width=0.20\textwidth]{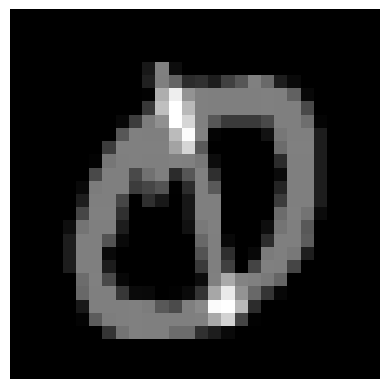}
  & \includegraphics[width=0.20\textwidth]{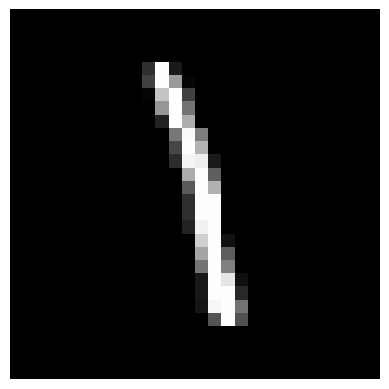}
  & \includegraphics[width=0.20\textwidth]{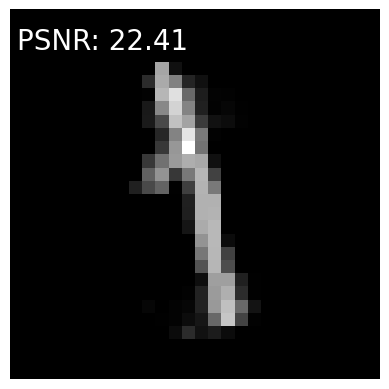}
  & \includegraphics[width=0.20\textwidth]{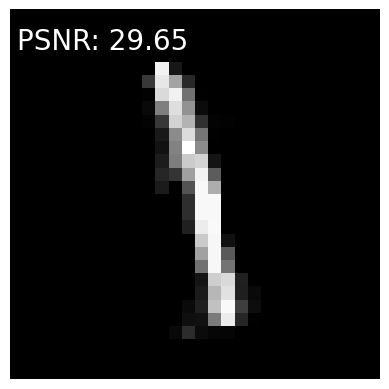}
  & \includegraphics[width=0.20\textwidth]{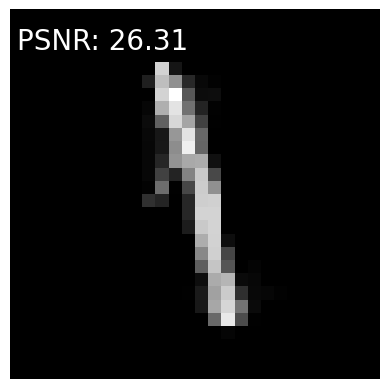} \\
  & \includegraphics[width=0.20\textwidth]{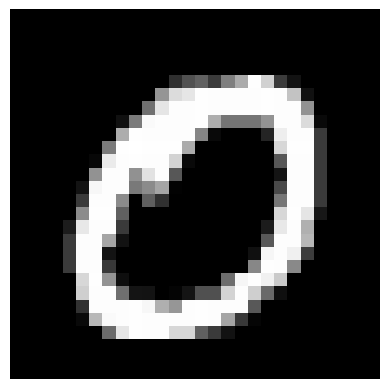} 
  & \includegraphics[width=0.20\textwidth]{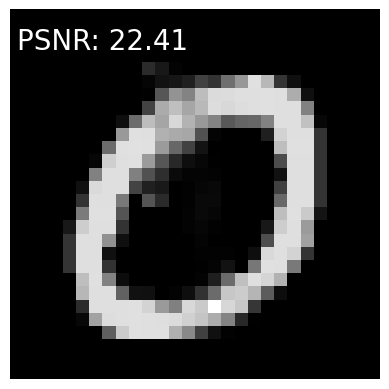} 
  & \includegraphics[width=0.20\textwidth]{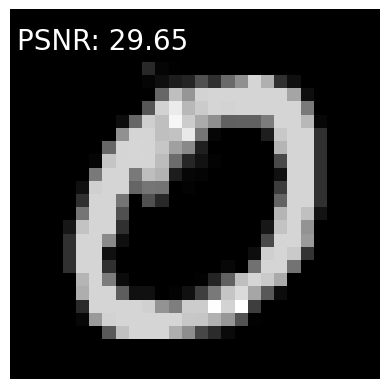} 
  & \includegraphics[width=0.20\textwidth]{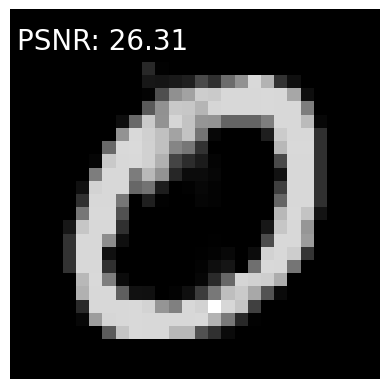} 
\end{tabular}
\centering
\caption{Example separation on test data for bases trained in data rich strong supervised setting with $d = 96$ and $\tau_A = 0.1$. 
All images are plotted independently so that the brightest pixel corresponds with the largest pixel value. 
The given PSNR value is the PSNR between the reconstruction and the true source data.
All separations carry some artifacts from the Wiener-filtering around the areas where the mixed images overlap. 
Standard NMF performs especially poorly and the separated images have some clear artifacts. We see this to a much smaller degree
for ANMF and DNMF, and they are qualitatively similar.}
\label{fig:data_rich_imgs}
\end{figure*}

An example of the separation results for a mixed image is shown in Figure \ref{fig:data_rich_imgs}.
We see that ANMF and DNMF appear to be better at learning which features belong to the different images.
While standard NMF only learns a set of features that can be used to reconstruct the images, ANMF and DNMF also learn
what features do not belong to that class of images.
The result of this is that these methods have less tendency to have features of
one source appear in another source, though this comes at the cost of losing some reconstruction accuracy for the relevant data.
Depending on the application, this ability to properly discern what feature belong to which source can be of higher importance than the overall reconstruction quality.

\subsubsection*{Experiment 2: Sparse data setting}

We now investigate the behavior of the methods in a more realistic, sparser data setting. 
We set the amount of strong supervised data to $N_\text{sup} = 250$
and the amount of weak supervised data for each source to $N_i  = 500$.
This emulates a realistic setting where obtaining strong supervised data is more difficult
than obtaining weak supervision data.
We also generate $N_\text{test} = 1000$ test data, which is not available during training, but will be used to compare the models.
The goal is to investigate how to best utilize this data to fit NMF bases.
We will attempt this by doing hyperparameter tuning as described in section~\ref{sec:hyperparameter}.

The parameters that need to be tuned for the different methods are shown in Table~\ref{tab:tune}.
We note that ANMF, and particularly D+ANMF, require more tuned parameters than the other methods, specifically the parameters $\tau_A$ and $\tau_S$.
\begin{table}[h]
\begin{tabular}{c|ccccc}
  \textbf{Tuning parameters}  & \textbf{ENMF} & \textbf{NMF} & \textbf{DNMF} & \textbf{ANMF} & \textbf{D+ANMF}  \\
  \hline
  $d$                         & \cmark  & \cmark & \cmark & \cmark & \cmark \\
  Initialization              & \cmark  & \cmark & \cmark & \cmark & \cmark \\
  $\mu_H$                     & \cmark  & \cmark & \cmark & \cmark & \cmark \\
  Test epochs                 & \cmark  & \cmark & \cmark & \cmark & \cmark \\
  $\mu_W$                     &         & \cmark & \cmark & \cmark & \cmark \\
  Training epochs             &         & \cmark & \cmark & \cmark & \cmark \\
  Batch sizes                 &         & \cmark & \cmark & \cmark & \cmark \\
  $\tau_A$                    &         &        &        & \cmark & \cmark \\
  $\omega$                    &         &        &        & \cmark & \cmark \\
  $\tau_S$                    &         &        &        &        & \cmark
\end{tabular}
\caption{Parameters that need to be tuned for different versions of NMF. Most of these parameters can also be tuned separately for each source.}
\label{tab:tune}
\end{table}

We choose not to tune $d$, as we saw in figure~\ref{fig:data_rich} and~\ref{fig:data_rich_tau} that 
results tend to improve with $d$ at the cost of computation speed and storage. We ideally
want the basis $W$ to include as few basis vectors as possible, and we choose $d=64$ for all experiments.

We also test for more classes of digits.
Images of ``one'' digits are most suited for NMF based methods, in the sense
that NMF bases trained on this digit do well in the reconstruction.
Therefore, we perform the experiment nine times, each time with a ``one'' digit mixed with 
a different digit. 
Because the overall results depend on which digits are mixed,
we chose to report how the methods perform in comparison to standard NMF. We therefore
report the difference in median PSNR between the desired
method and standard NMF, denoted $\Delta \text{Median PSNR}$.

Given enough data and a sufficiently good parameter search, D+ANMF should always outperform or do equally well as the other methods, as it is a superset of all of them.
This might not be the case, though, if the models overfit the data, or if the parameter search is too coarse.

\begin{figure}[!htb]
  \centering
  \includegraphics[width = 0.8\textwidth]{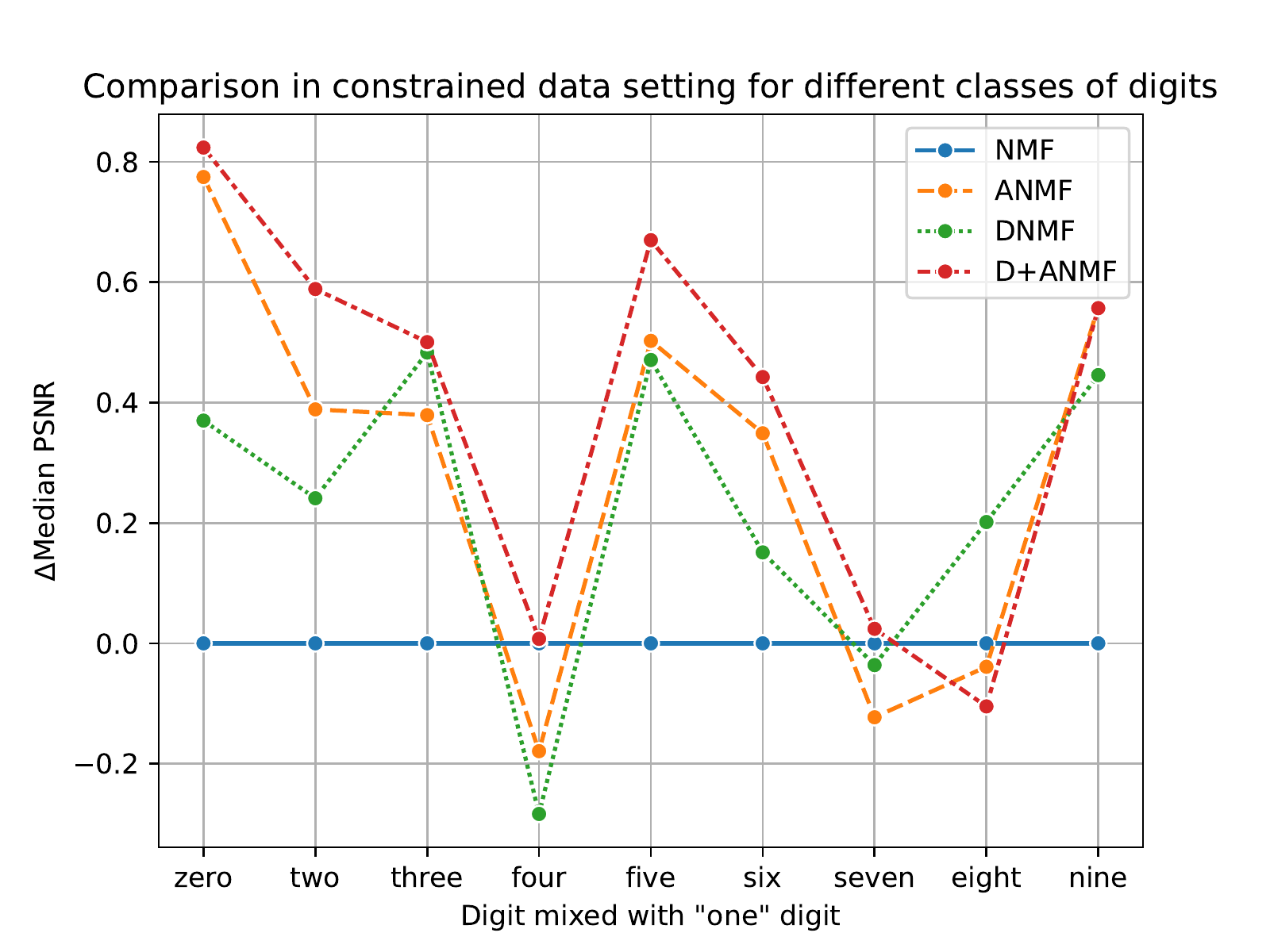}
  \caption{Results from tuning experiments with different digits in the low strong supervision data setting.
  The $y$-axis is the difference in median PSNR between the method and standard NMF.
  The digits on the $x$-axis illustrate which digit was mixed with ``one'' digits when synthetically generating data.
  We observe that when mixing with ``four'', ``seven'' and ``eight'' digits, all methods perform very similarly.
  For the other digits, we see that D+ANMF performs best, closely followed by ANMF and DNMF. }
  \label{fig:data_poor}
\end{figure}

The distributions used for the hyperparameters in the random search implementation can be found in the supplementary material.\footnote{See \url{https://github.com/martilud/ANMF}.}
For each fit we try $15$ different randomly sampled parameters. The results are shown in Figure~\ref{fig:data_poor}.
The results seem to indicate that for some digits, or some train--test splits, there is little performance
gain from using more complex methods than standard NMF.
One reason for that is the presence of an upper limit to what can be learned
by non-negative linear bases, especially when the amount of data is low. 
For the other digits, we observe that D+ANMF performs best, closely followed by ANMF and DNMF.
This indicates that methods that can utilize all available data can outperform methods
that are restricted by the availability of strong supervision data, at the cost of more computation time and
more parameters that need to be tuned.

\subsection{Audio data}

We will now perform a small speech denoising experiment to further exemplify the usage of 
our proposed methods.
For that, we follow the standard approach for using NMF for audio source separation \cite{fevotte2018single,weninger2014discriminative}.
Also, we only consider the semi-supervised case,
where we have clean speech recordings and noisy speech recordings with different speech, but no clean recordings of noise.

\subsubsection*{Data}
In order to make the experiment reproducible, we only use open-source data.
For speech, we use the LibriSpeech dataset \cite{7178964}, which contains $1000$ hours of public domain recordings
of English audiobooks recorded at $16$kHz. We will only use a tiny part of the total corpus.
Specifically, we will use recordings of a single female speaker over two different audiobooks,
which corresponds to $8$ minutes of speech spread over $42$ audio clips that are about $11$ seconds each.
For noise, we use the WHAM! dataset \cite{Wichern2019WHAM}. This dataset contains recordings of 
various urban locations like restaurants, caf\'es, bars, and parks, also recorded at $16$kHz.
This is a relatively challenging noise dataset, with highly non-stationary and varied noise signals.
We split the dataset into training and testing, where the train dataset contains half
of the speech data, and the test set contains half of the speech data with added noise.
We mix the noise additively at different input Signal-to-Noise ratios (SNR).

\subsubsection*{Feature extraction}

We extract features by taking the Short-Time Fourier Transform (STFT), and take the amplitude
to obtain the spectrum of the audio. We then attempt to separate on the spectrum, and use Wiener-filtering
to reintroduce the phase before applying the Inverse STFT (ISTFT). 
The librosa package is used to implement the STFT \cite{mcfee2015librosa}. 
For the STFT we use $512$ samples for the FFT of each window, which corresponds
to a relatively low latency window length of $32$ ms.

\subsubsection*{Method}

For speech denoising applications, we are only concerned with reconstructing speech, which means
we can apply NMF in two ways. The first way is to 
train a basis for the known signal using NMF or ANMF, then solve equation \eqref{eq:semi}
to obtain a basis for the unknown noisy signal, and finally use this to separate the signal.

The second way is to only learn a basis for the known signal and then project the mixed
signal onto this basis. This approach is only suitable for low-noise problems where 
the noise is sufficiently independent of the speech signal. This approach has the advantage
that we do not need to learn a basis for the noise, which can be computationally expensive.
We call this projection approach P-NMF and P-ANMF. 
To measure quality, we use SI-SDR \cite{le2019sdr}, which we apply to the individual audio clips. 

\subsubsection*{Results}

Results for the audio data can be found at \url{https://github.com/martilud/ANMF}.
In contrast to multi-speaker denoising applications with larger datasets, we have no need for sparsity,
and we can select the number of basis vectors relatively low.  
Specifically, we select the number of basis vectors as $d = 64$ for both the speech and noise sources, and the sparsity parameters $\mu_W = \mu_H = 10^{-10}$, mainly to stabilize the numerical algorithms.
We observe that selecting larger sparsity parameters can lead to drastically worse performance, both quantitatively and qualitatively.

For ANMF, we select $\tau_A = 0.5$, and use the parameter $\beta$ as described in equation \eqref{eq:beta}. To calculate $\beta$, we use the exact
weights that were used for mixing the noisy signals. The results are shown in Table \ref{tab:audio}, where we see that ANMF outperforms NMF for all noise levels.
We also observe that P-ANMF more heavily outperforms P-NMF, and even outperforms NMF, which is remarkable as P-ANMF does not fit a basis to the unseen source.
We also observe in the experiments that there is still much room for tuning $\tau_A$ and $\beta$ for the specific noise levels for ANMF.
\begin{table}[h]
  \centering
  \begin{tabular}{l|cccccc}
  SI-SDR [dB] & \multicolumn{6}{c}{Input SNR [dB]} \\
   & -6 & -3 & 0 & 3 & 6 & 9 \\
  \hline 
  NMF & 0.84 & 3.52 & 6.04 & 8.43 & 10.48 & 12.16 \\
  ANMF & \textbf{1.39} & \textbf{4.23} & \textbf{6.98} & \textbf{9.18} & \textbf{11.35} & \textbf{13.09} \\
  \hline 
  P-NMF & -1.85 & 1.10 & 4.02 & 6.89 & 9.64 & 12.19 \\
  P-ANMF & 0.07 & 2.96 & 5.79 & 8.38 & 10.72 & 12.77 \\
  \end{tabular}
  \caption{Mean SI-SDR for a speech denoising experiment for different noise levels. Because the noise is purely additive,
  the SI-SDR of the noisy signals are very close to the input SNR.}
  \label{tab:audio}
\end{table}

Qualitatively we note that the denoised signals produced by ANMF are of higher or equal quality. A feature of ANMF is that it consistently
removes more stationary (low-frequency) noise from the signal, as this is the part of the signal that can most safely be removed without noticeably damaging the quality of the speech signal.
In contrast, the projection-based denoised signals, although not much worse quantitatively, have clear artifacts, making them much less preferable.
Finally, we see that all methods perform worse at removing non-stationary noise, like background music or sharp sounds from moving objects.
In order to be able to treat this type of noise with NMF-based methods, we suspect one would need more basis vectors to fully model the complexity of the noise, as well as more data to properly fit the bases.
Another approach would be to use generative models that are better capable at modeling non-stationary noise data.

The conclusion is that ANMF can learn compact bases that can be used for single speaker speech denoising in a semi-supervised data setting. 
We believe ANMF would also be suitable for more large scale multi-speaker denoising and other audio applications, but this would require further investigation.

\section{Further Work}
Several potential avenues for future research are worth exploring,
including investigating adversarial training of more complex generative methods, incorporating transfer learning techniques,
and identifying better approaches for parameter tuning.
We believe that adversarial training can be a valuable tool for source separation problems in weak supervision data settings.
The main challenge is how to most efficiently utilize the trained models and data available to achieve good and robust results.
While NMF-based approaches can be powerful because of their simplicity and implicit regularization, a more complex feature extraction and/or more complex generative 
models are most likely needed to achieve better results, at least for non-stationary signals.

\section{Conclusion}
In this article we have investigated adversarial generative training for single channel source separation. In particular, we have developed a variant of Non-Negative 
Matrix Factorization that we called Adversarial Generative Non-Negative Matrix Factorization (ANMF). 
We have discussed how to utilize weak and strong supervision data for training adversarial generative functions for source separation problems.
Moreover, we have introduced a numerical algorithm for fitting ANMF with stochastic updates.
We have seen in the numerical
experiments that ANMF outperforms existing NMF methods 
for both image and audio source separation problems,
including methods that make use of strong supervision data.

\end{document}